\documentclass[twocolumn,showpacs,preprintnumbers,amsmath,amssymb]{revtex4}
\usepackage{graphicx}% Include figure files
\usepackage{dcolumn}% Align table columns on decimal point
\usepackage{bm}% bold math
\begin{document}
%\draft

\title{Linear Response and the Thomas-Fermi Approximation in Undoped Graphene}

\author{L. Brey}
% J. J. Palacios$^{1,2}$. }
\affiliation{Instituto de Ciencia de Materiales de Madrid
(CSIC), Cantoblanco 28049, Spain}
%\\ (2)Departamento de F\'{\i}sica Aplicada, Universidad de Alicante, San Vicente del Raspeig, E-03690 Alicante
%Spain}
\author{H.A. Fertig}
\affiliation{Department of Physics, Indiana University, Bloomington, IN 47405}

\date{\today}

\begin{abstract}

We analyze the range of validity of Thomas Fermi theory for
describing charge density modulations induced by external potentials
in neutral graphene. We compare exact results obtained from a
tight-binding calculation with those of linear response theory and
the Thomas Fermi approximation. For experimentally interesting
ranges of size and density amplitudes (electron densities less than
$\sim 10 ^{11} cm ^{-2}$, and spatial length scales below $\sim 20
nm$), linear response is significantly more accurate than Thomas
Fermi theory.

\end{abstract}  \maketitle

\section{Introduction}
The realization of single flakes of graphene
 -- atomically thin layers of carbon atoms packed in a honeycomb lattice -- has
made possible the experimental study of two-dimensional massless Dirac
fermions \cite{Novoselov_2004,Novoselov_2005,Zhang_2005}. Graphene is
a gapless semiconductor in which the conduction and valence bands
touch at two points -- Dirac points -- in the Brillouin
zone \cite{Castro_Neto_RMP}. Near either of these points the electronic states
are described by a massless Dirac equation, with eigenstates which are spinors
due to the two-point basis needed to describe the honeycomb lattice \cite{Ando_2005}.
The effective spinors of the wavefunctions are either parallel or antiparallel
to the momentum, so that the states are chiral.

For undoped graphene there is one electron per carbon atom,
and the system ideally should be everywhere charge neutral.
In practice this is known not to be the case.
Recent imaging experiments \cite{Martin_2007} have demonstrated the
existence of electron and hole puddles of densities $\sim 10^{10}
-10^{11}$ cm$^{-2}$ in the vicinity of the neutrality point. The
existence of these charge puddles could be related to the existence
of mechanical ripples also observed in graphene
sheets \cite{Meyer_2007,Stolyarova_2007,Ishigami_2007}, which can cause
modulation of the electronic
charge \cite{Brey_Palacios_08,Guinea_ripples_08}, or to unintentional
charged impurities in the substrate \cite{Hwang_2007,Nomura_2006,Ando_2006},
which can also generate electron-hole
puddles \cite{Rossi_2008,Polini_2008,Fogler_2008}.  The spatial
correlation length of these puddles is of the order of 10$nm$.

Local density inhomogeneities can also be induced in graphene using
miniature gates. In this way graphene $p$-$n$ junctions have been
experimentally realized \cite{Huard_2007,Ozyilmaz_2007,Williams_2007}. Recent
advances in the quality of graphene have made possible the
fabrication of ballistic circuits with electrically controlled $p$-$n$
junctions\cite{Young_2008,Stander_2008}.

The physical properties of graphene with such electronic inhomogeneities
depend strongly on the size and amplitude of  charge
modulation induced by external potentials. It is therefore important to
understand how the ground state
charge in graphene is distributed in their presence.
Large inhomogeneous graphene systems have been studied theoretically
using the Thomas Fermi (TF) approximation, which, as we discuss below,
treats the
kinetic energy in a local density approximation \cite{Zhang_2008}.
Rossi and {Das Sarma} used a TF
approximation with Hartree and exchange effects included to study
the ground state of neutral graphene in the presence of charged
impurities \cite{Rossi_2008}. A more rigorous quantum mechanical
treatment of the kinetic energy is possible, but its use
limits considerably the system sizes
which in practice can be studied \cite{Polini_2008}.

As we will show below, because of the crossing of the chiral
electron and hole bands at the Dirac point, the TF approximation
does not correctly capture the charge response of neutral graphene
to an external potential in many interesting situations.
The purpose of  this work is to
analyze the range of validity of the TF theory near the Dirac point.
We use a microscopic tight-binding calculation to compute the
response of neutral graphene to electrostatic potentials, and
compare these exact results both with linear response and with the TF
approximation. We will demonstrate that for experimentally interesting \cite{Martin_2007}
ranges of sizes and amplitudes
(electron densities $\sim 10 ^{11} cm ^{-2}$ and spatial correlations $\sim
20 nm$), simple linear response results match exact results quite well, while results of the TF
approach are much poorer.
The failure of the TF approximation is related to the non-local
character of the density response, and we shall see that
a kinetic energy functional which
correctly captures the linear response of neutral graphene to
external electrostatic perturbations has a highly non-local nature.

\section{Thomas Fermi functional for the kinetic energy.}
\subsection{Formal Considerations}

Following
Hohenberg and Kohn \cite{Hohenberg_1964}, the total energy of
the {noninteracting} system, $E$, may be written in terms of a kinetic energy functional $T[n({\bf r})]$ of the
electron density $n({\bf r})$,
\begin{equation}
E[n({\bf r})]=\int T[n({\bf r})] d{\bf r} + \int V({\bf r}) n ({\bf
r})d{\bf r} \, \, \, \, . \label{func_total}
\end{equation}
Here $V({\bf r})$ is the one-body external potential in which the
particles move, and the density is defined with respect to the density
of  electrons in neutral graphene.  The effect of electron-electron interactions in
a Hartree approximation will be considered below
in Section III.

The TF theory assumes that the functional $T[n({\bf r})]$ is a
local function of the density, and the form of the functional is chosen
such that for a uniform potential $V$ the minimization of Eq. \ref{func_total}
recovers the kinetic energy of a homogeneous system. For
the case of Dirac fermions, the Thomas-Fermi kinetic energy functional is
\begin{equation}
T[n]= \hbar v_F \, \frac { 2 \sqrt{\pi}}{3} \,  \textrm{sgn} [n({\bf
r})] \,  |n ({\bf r})|^{3/2} \label{TF_funct} \, ,
\end{equation}
where $v_F$ is the Fermi velocity of the carriers near the Dirac
point. The minimization with respect to the density must be
carried out subject to the normalization constraint
\begin{equation}
\frac 1 S \int n({\bf r}) d {\bf r} = n_0 \, \, \, ,
\end{equation}
where $n_0$ is the average electron density measured relative to
that of undoped graphene, and $S$ is the sample area. Minimizing
Eq.\ref{func_total} yields the relation
\begin{equation}
n^{TF} ({\bf r}) = \frac 1 {\hbar ^2 v_F ^2 \pi } \, \textrm{sgn}
\left ( \mu _0  - V({\bf r}) \right ) \left( \mu _0 - V({\bf r})
\right ) ^2, \label{dens_TF}
\end{equation}
where $\mu _0 = \hbar v_F \sqrt{\pi n_0}$  is the Fermi energy of the
corresponding homogeneous system.  Defining $k_{max} ({\bf r}) =
\sqrt{\pi |n ^{TF} ({\bf r})|}$, we find that the carriers
have higher kinetic energy where the potential energy is
lower, and vice-versa. Eq. \ref{dens_TF} van be viewed as the relation
between the local maximum momentum $k_{max}({\bf r})$ and the external
potential $V({\bf r})$ obtained from a classical equation of motion.

An interesting and important consistency check of the
TF approximation \cite{Hohenberg_1964} is that the
response function of the system should be directly related to the second
functional derivative of the energy. For a non-interacting system
this takes the form
\begin{equation}
\mathcal{F} \left ( \left . \frac { { \delta ^2 T [ n ({\bf r})] }}{
\delta n ({\bf r}_1) \delta n ({\bf r}_2)}   \right | _{n_0}  \right
) = - \frac 1 {\chi _{Lin} (q,n_0)},
\label{susceptibility}
\end{equation}
where $\mathcal{F}$ indicates the Fourier transform, and $\chi _{Lin}
(q,n_0)$ is the wavevector dependent static Lindhard susceptibility
of the uniform system at density $n_0$.

In graphene the Lindhard static susceptibility in the long
wavelength limit has the form\cite{Hwang_2006,Wunsch_2006,Ando_2006}
\begin{eqnarray}
\chi _{Lin} (q,n_0) \!&  \! = \! & \! - \frac 2 {\sqrt{\pi}} \frac
{|n_0|^{1/2}}{\hbar v_F} \, \, \, \,  \textrm{for} \, \, \, \, q \! < \! \sqrt{\pi n_0}, \nonumber \\
\chi _{Lin} (q,n_0) \!& \!  = \!  & \! - \frac q {4 \hbar v_F} \, \,
\, \, \, \, \, \, \, \, \, \, \, \, \, \, \, \textrm{for} \,\,
n_0=0. \label{dirac_response}
\end{eqnarray}
The TF kinetic energy functional, Eq.\ref{TF_funct}, correctly recovers
the response function for doped
graphene in the long wavelength limit, but it fails to describe the non-interacting
compressibility at the neutrality point. In fact, the TF
approximation predicts vanishing linear response
at the Dirac point. This failure is in agreement with the general assumption
of the TF theory that $|\nabla n ({\bf
r})|/[n({\bf r}) k_{max}({\bf r})] \ll 1$, which cannot be satisfied near
charge neutrality.  {Moreover, as we discuss below, the response in the
second of Eqs. \ref{dirac_response} is inherently non-local, suggesting that
the TF approximation must break down near charge neutrality.}

\subsection{Numerical results.}

\begin{figure}
  \includegraphics[clip,width=8cm]{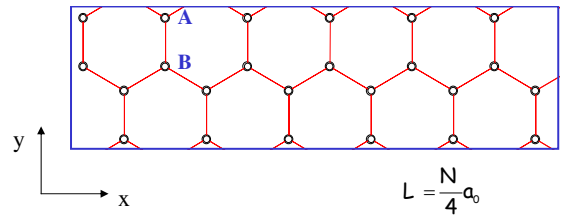}
  \caption{($Color$ $online$)Unit cell used in the calculations. The unit cell contains $N$ atoms
  and the length of the unit cell is $L= N/4 a_0$. The external potential only depends on $x$ and
  in this geometry atoms $A$ and $B$ with the same $x$ coordinate have the same
  charge \cite{Brey_2006a,Brey_2006b}. $a_0$ is the lattice parameter of the triangular lattice.}
   \label{Figure1}
\end{figure}

In order to quantify the effects of the failure of the TF approximation
to correctly describe the linear response of undoped graphene,
we
numerically compute the electron density of a net neutral graphene system in an external
potential, and compare the results with the TF approximation and with linear
response results. We use a simple tight-binding Hamiltonian with nearest neighbor hopping, of
the form
\begin{equation}
H=t \sum _{<i,j>} C^+ _i C_j + \sum _i V_i C^+ _i C_i \, \, \, ,
\label{Hamil_tb}
\end{equation}
where $C_i$ annihilates an electron at site ${\bf R} _i$ of the
graphene lattice, $t=\frac 2
{\sqrt{3}}\frac {v_F}{a_0}$ is the hopping paramenter, $a_0$ is the lattice parameter of
the triangular lattice, and  $V_i$ represents the external
potential at site ${\bf R }_i$. We perform the calculations in a
unit cell illustrated in Fig. \ref{Figure1}, using periodic
boundary conditions in both the $x$ and $y$ directions. The external
potential  and the induced charge depend only on the $x$
coordinate. In the unit cell represented in Fig.\ref{Figure1} atoms
on both sublattices  experience the same external potential, so
there is no out-of-phase response from atoms on
different sublattices\cite{Brey_2007b}.

\begin{figure}
  \includegraphics[clip,width=8cm]{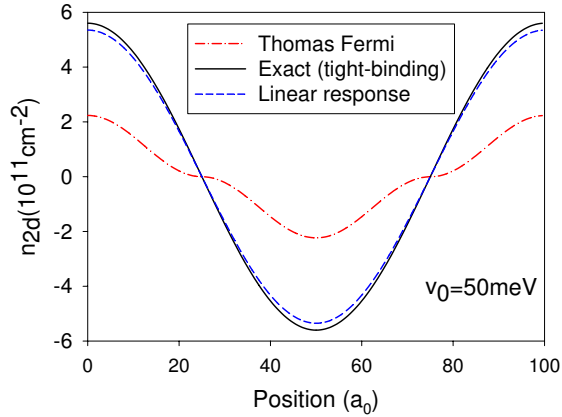}
  \caption{($Color$ $online$)  Density profiles
  obtained with different approximations for  perturbation amplitude $V_0=50meV$ and period 100$a_0$. For
  exact calculations, $t=2.8eV$ and $a_0=2.46\AA$.
Solid line is the exact result, dashed line indicates linear response
result,
   and
  dash-dotted line is result of Thomas Fermi approximation.
  }
   \label{Figure2}
\end{figure}

We study the response of the system to the potential
\begin{equation}
V_i = V_0 \cos (G  X_i) \,\label{V_ext}
\end{equation}
where $X_i$ is the $x$-component of the position of the carbon
atoms, and $V_0$ is the amplitude of the perturbation.
Fig.\ref{Figure2} illustrates a typical result, the electron density induced by a
potential of amplitude $V_0=50meV$ and period 100$a_0$.
Also plotted are
the density as obtained in linear response, $n^{Lin}(G)=\chi _{Lin}
(G,0) V_0$, and from the TF approximation, Eq. \ref{dens_TF}. The
density induced by this potential is of the same order as the
densities of electron and hole puddles observed experimentally. Note
that the linear response reproduces the exact
result rather faithfully, whereas for this potential the TF approximation
underestimates the response.
Moreover, the TF approximation
displays plateau-like features  when passing through zero density, which
are an artifact of the approximation \cite{Zhang_2008}; they appear because TF theory
grossly underestimates the ability of the system to screen when
the local chemical potential is
near the Dirac point. The plateaus may be understood more formally by substituting
the perturbation Eq. \ref{V_ext} into Eq. \ref{dens_TF} and expanding in
harmonics, to obtain
\begin{equation}
n ^{TF} (x)=- \frac { V_0 ^2 \textrm{sgn} (V_0)} {\hbar ^2 v_F ^2
\pi } \frac 8 {3 \pi} \left ( \cos G x+ \frac 1 5 \cos3Gx+... \right
) \, \, \, .
\end{equation}
The large  $\cos3Gx$ harmonic leads to the plateau-like behavior
when crossing the Dirac point. %In the exact solution the third
%harmonic shows proportional to $V_0 ^3$ and only is apparent in the
%limit of very small $G$ or very large $V_0$.

\begin{figure}
  \includegraphics[clip,width=8cm]{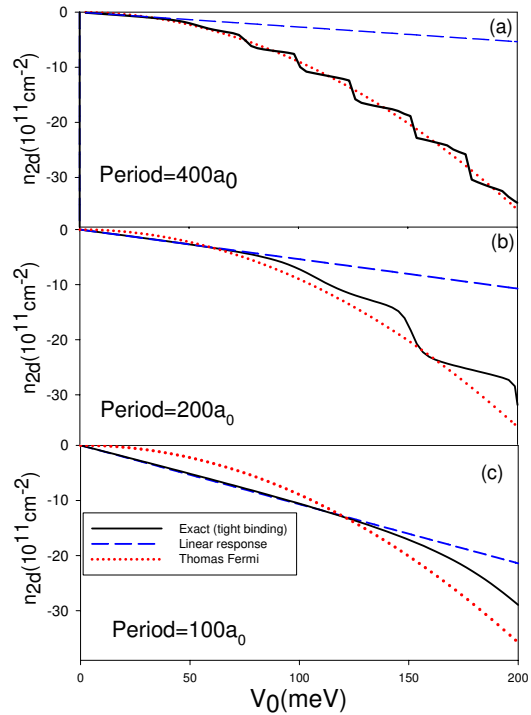}
  \caption{($Color$ $online$) Maximum  induced  electron
  density (density at $x$=0) as function of the amplitude of the external potential. The external potential has the form
  $V(x)=V_0 \cos {Gx}$. (a), (b) and (c) correspond to values $Ga_0=\pi/200$, $Ga_0=\pi/100$ and
  $Ga_0=\pi/50$ respectively.  Continuous lines are the exact results, dashed lines are the linear response results,
   and
  dash-dotted lines are the results obtained in the Thomas Fermi approximation.  In the
  calculations we use the values $t=2.8eV$ and $a_0=2.46\AA$.}
   \label{Figure3}
\end{figure}

In Fig. \ref{Figure3} we compare  the maximum electron density at
$x=0$,  obtained both from the exact calculation, and in the two
different approximation schemes, as a function of $V_0$, for
different periods of the external potential. For small periods and
small  $V_0$, the linear response results follow the exact results
rather closely. TF theory by contrast underestimates the response of
the system. For small enough $V_0$ and large  $G$, linear response
is able to properly capture the non-local nature of screening in
this system. For large wavelengths and  external potentials
non-linear contributions to the response become important, and may
be captured by the TF approximation in any average way (Fig.
\ref{Figure3}(a).) From the numerical results we  estimate that, in
the absence of electron-electron interactions, linear response is
more reliable than TF when $n^{Lin} > n^{TF}$. For large
perturbations the exact density response oscillates around the TF
result. These oscillations are induced by  zero modes created by the
external potential in graphene\cite{Brey_2009a}, which cannot be
captured by a local approach such as the TF approximation.

For the charge density modulation amplitudes observed
experimentally, $\sim 10^{11} cm ^{-2}$, the length scale for which
linear response is more reliable than the TF approximation is larger than the size
of the observed electron-hole puddles \cite{Martin_2007}. Furthermore, from the geometry of the multiple
gated graphene devices in Refs. \onlinecite{Young_2008} and \onlinecite{Stander_2008}, we find that
the width of the depletion regions in the $p$-$n$
junctions \cite{Zhang_2008} are also smaller than the length scale
where linear response is applicable. More generally, our results indicate that
\emph{for density modulations up to $10^{12} cm ^{-2}$ on length
scales up to 20$nm$, linear response results are significantly more
accurate than those of the TF approximation.}
This conclusion agrees  with results presented in Fig. 2 of
Ref. \onlinecite{Polini_2008}, where the authors find results
which are consistent at
semi-quantitative level with a linear screening theory.

\section{Hartree interaction}

\subsection{Formulation in Terms of Linear Response}
Any modulation of electric charge produces a change in the
energy associated with the repulsion between electrons.
If one is interested in the long-wavelength static response of
the charge density to a potential inducing such a modulation,
the most
important effects of the electron-electron interaction can be
captured by the Hartree energy.  This may be written in the form
\begin{equation}
E_H = \frac 1 2 \frac {e ^2} {\varepsilon} \int d {\bf r} \int d
{\bf r}' \frac {n({\bf r}) n ({\bf r}')}{ |{\bf r} - {\bf r}'|}
\label{Hartree} \, \, \, ,
\end{equation} where $\varepsilon$ is the average background
dielectric constant. The strength of the Coulomb interaction is
given by the dimensionless parameter
\begin{equation}
\alpha= \frac {e ^2}{ \hbar v_F \varepsilon} \,\, \,.
\end{equation}
For graphene on a conventional SiO$_2$ substrate, $\varepsilon
\approx 2.5$ and $\alpha \approx 0.9$.  For substrates with larger
$\varepsilon$ such as HfO$_2$ or liquid water, the values of $\alpha$
can be much smaller.
We note that in principle one may improve upon the Hartree approximation by
including
exchange correlation effects,
but for chiral Dirac fermions these
appear to be rather small \cite{Polini_2008}.

\begin{figure}
  \includegraphics[clip,width=8cm]{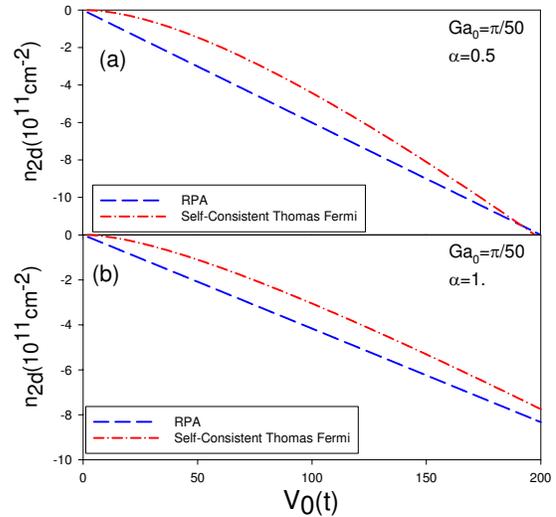}
  \caption{($Color$ $online$) Maximum induced electron density as a function of the amplitude of an external potential of period
100$a_0$, in the Hartree approximation.
  (a) corresponds to a weak electron-electron
  interaction, $\alpha$=0.5
  and (b) to a stronger one, $\alpha$=1.
   Dashed lines represent RPA results; dash-dotted lines are results obtained by minimizing
   an energy functional containing both the TF approximation to the kinetic energy and
   the Hartree form of the Coulomb interaction.
 }
   \label{Figure4}
\end{figure}

Since we will
consider perturbations with  amplitudes and periods such that the
exact non-interacting result coincides nearly perfectly
with that of linear response, we expect that
the inclusion of the Hartree term leads only
to linear screening of the external potential. In this case the
induced charge coincides with that obtained in the Random Phase
Approximation (RPA). In reciprocal space this means
\begin{equation}
n ^{exact} (G) \simeq n ^{RPA} = \frac {\chi _{Lin} (G,0)}{1-v_G \,
\chi _{Lin} (G,0)} V_0,
\end{equation}
where $v_q=\frac{ 2 \pi e ^2}{\varepsilon q }$ is the two
dimensional Fourier transform of the Coulomb interaction.

In the TF approximation the electron density is
obtained by minimizing the kinetic functional, Eq. \ref{func_total}, together
with the Hartree energy Eq. \ref{Hartree} with respect to the density. In Fig. \ref{Figure4} we
compare the spatial maximum electron density
obtained in the RPA to the
TF approximation as a function of the amplitude of the external
potential. The Hartree interaction screens the external potential,
so that the induced charge density decreases with increasing
electron-electron interaction parameter $\alpha$. As in the non-interacting
case we see that the TF approximation underestimates the response at
small $V_0$. For physically relevant values of
$\alpha$, we see that the TF approximation is not quantitatively reliable in describing the response
of neutral graphene to external potentials that generate density fluctuations
of magnitude $10 ^{12} cm ^{-2}$ or below, within length scales of about 20$nm$.

\subsection{Electric fields in a $p$-$n$ junction}
Ballistic transport in  graphene $p$-$n$ junctions is due to Klein
tunneling of the massless electrons. Cheianov and Fal´ko
\cite{Cheianov_2006a} showed that the ballistic resistance per unit
width of a graphene $p$-$n$ junction is $R=\frac {\pi} 2 \frac h
{e^2}\sqrt{\frac {\hbar v_F }{ e E}}$, where $E$ is the assumed
uniform electric field at the junction. Note the resistance
decreases as the electric field at the interface decreases. This
electric field depends on the screening properties of graphene near
the Dirac point. Zhang and Fogler \cite{Zhang_2008} proposed  that
the electric field in the depletion region separating the electron and hole regions
is enhanced due to the limited screening capacity of Dirac
quasiparticles.

\begin{figure}
  \includegraphics[clip,width=8cm]{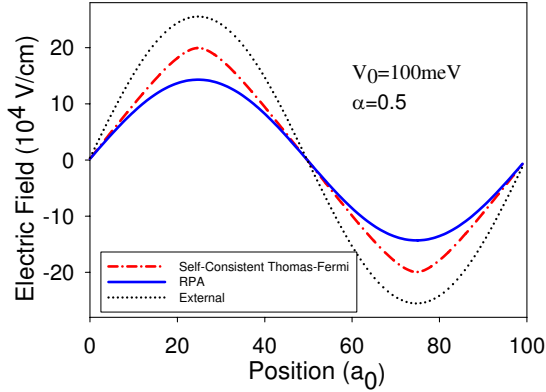}
  \caption{($Color$ $online$) Electric field as a function
  of position for a periodic external potential of amplitude $V_0$=0.1eV and
  period 100$a_0$.  Coupling constant is taken to be $\alpha=0.5$.}
   \label{Figure5}
\end{figure}

In order to study the difference in computed values of $F$, the
electric field in the depletion region, using the TF approximation
and linear response theory,  we have calculated the electric field
for a cosine-shaped external potential Eq.\ref{V_ext}. This
potential creates periodic electron and hole regions separated by
$p$-$n$ interfaces. In Fig.\ref{Figure5} we plot the electric field
as a function of position, as obtained in the TF approximation, and
in the RPA (the latter being essentially an exact solution of the
Hartree approximation.) For comparison we also plot the applied
electric field, $E^{ext} =- G V_0 \sin{ G x}$. The results presented
are for $\alpha$=0.5. The $p$-$n$ and $n$-$p$ interfaces are located
at $x$=25$a_0$ and $x$=75$a_0$, respectively. At these points the
values of the electric field are maximal. In the linear calculation
the electric field $F$ can be calculated analytically, yielding the
result $F=V_0 \, G /(1+\pi / 2 \alpha )$, so that the external
electric field is reduced by a factor $(1+\pi / 2 \alpha )$ by the
screening. In the TF approximation a numerical minimization is
required to obtain $F$. In the range of validity of the linear
approximation we find that the TF approach predicts much weaker
screening of the external field than the RPA.

\begin{figure}
  \includegraphics[clip,width=8cm]{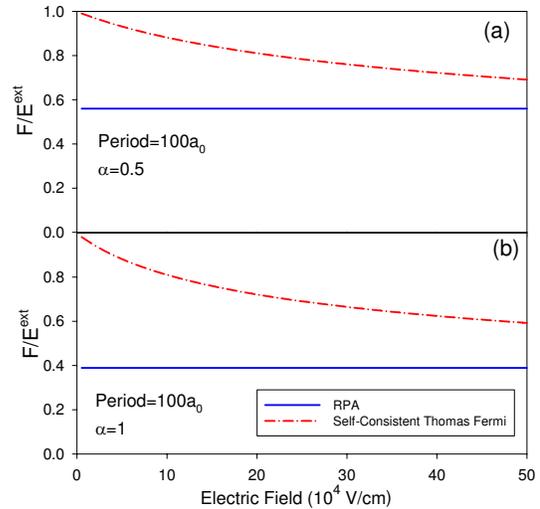}
  \caption{($Color$ $online$) Electric field at the $p$-$n$ junction centers as a function of the external field.
  (a) corresponds to a weak electron-electron
  interaction, $\alpha$=0.5
  and (b) to a stronger one, $\alpha$=1.
   Continuous  lines corresponds to the RPA results. Dash-dotted lines are the results obtained
   in the self-consistent Thomas-Fermi approximation.
  }
   \label{Figure6}
\end{figure}

In Figs. \ref{Figure6}(a) and \ref{Figure6}(b) we plot the values of
the electric field at the $p$-$n$ junctions, normalized to the
external field, as a function of the applied electric field, for two
different values of $\alpha$.  The screened  electric
field at the interface obtained from the TF theory is larger than that obtained in
the linear response theory (RPA), as expected from the above results.
We see that
the TF approximation significantly overestimates the total electric field at the $p$-$n$
junction.

\section{Summary and Observations}

The Thomas Fermi approximation is relatively inaccurate for describing density
modulations for wavevectors that are not too small, and external
potentials which are not too large, in undoped
graphene. Quantitatively this region of failure of the TF approximation
appears to apply to the observed density fluctuations
of the electron-hole puddles that appear in the
single electron transistors spectroscopy.
It also appears to be problematic for estimating the electric field
in a graphene $p-n$ junction.
The reason for its failure is its inability to capture the
intrinsically non-local response of neutral
graphene.  We find that the application of linear response theory (RPA)
in this regime is far more quantitative.

It is interesting to note that one may adopt a non-local kinetic
energy functional to produce a correct result for Eq.
\ref{susceptibility}. This takes the form \cite{private_allan}
\begin{equation}
T^{linear} [ n({\bf r})]= \frac { \hbar v_F} {\pi} \int d {\bf r} \int d
{\bf r}' \, \, \frac {n({\bf r}) n ({\bf r}')}{ |{\bf r} - {\bf
r}'|}
\label{nonlocal}
\end{equation}
This kinetic energy functional is formally the same as the Hartree form of
the interaction energy, highlighting the marginal nature of $1/r$ Coulomb
interactions in undoped graphene \cite{Castro_Neto_RMP}.  Its long-range nature
strongly suggests the difficulties of a local approximation such as TF that
we find.

To improve upon the TF approximation one can formally compute first
order gradient corrections to the density using a WKB approximation
applied to the Green's function \cite{Pfirsch}.The result
\cite{herb_unpub}, however, has singular behavior near zero
momentum, and moreover depends locally on both the density and its
gradient, and so cannot produce corrections where the density is
maximum and where TF has significant errors.

Finally we note that Eq. \ref{nonlocal} may be used to develop a criterion
for which one expects the TF approximation to fail.  Using the result of the linear response density
in Eq. \ref{nonlocal} gives an estimate for the energy density expected from the non-local
contribution to the energy. Comparing this to the TF energy density (Eqs. \ref{TF_funct} and \ref{dens_TF}),
we expect to the latter to be larger if the TF approximation is to be valid.  This yields
the criterion $V_0/G > \eta \hbar v_F$, where $\eta$ is a geometric factor of order 1,
for which the TF kinetic energy dominates over non-local contributions to the energy.
Notice this means that, for fixed length scale $1/G$, the TF approximation will always
fail for sufficiently small potential scales $V_0$.

%\begin{figure}
%  \includegraphics[clip,width=8cm]{ripples-new.eps}
%  \caption{($Color$ $online$) Schematic view of graphene with periodic ripples.
%  $L$ denotes the lateral extent of the ripples and $h$ the out of plane displacements of the Carbon atoms.}
%   \label{ripples}
%\end{figure}

\section{Acknowledgments.}

We acknowledge useful discussions with F. Guinea, M. Polini,
E. Chac\'on, S. Das Sarma, and A.H. MacDonald.  We thank the Aspen Center for
Physics for hospitality where this research was initiated, and the KITP at UCSB
where it was completed.  This work was been
financially supported by MEC-Spain MAT2006-03741 and by the National Science
Foundation through grant No. DMR-0704033.

%\bibliography{mia}

\end{document}